\documentclass[]{aastex63}

\usepackage[utf8]{inputenc}
\usepackage{amsmath}
\usepackage{graphicx}
\usepackage{float}
\usepackage{xfrac}

\newcommand{\RomanNumeralCaps}[1]{\MakeUppercase{\romannumeral #1}}

\received{February 19, 2021}
\revised{April 26, 2021}
\accepted{April 29, 2021}
\submitjournal{AJ}

\shorttitle{Flare Induced Photospheric Velocity Diagnostics}
\shortauthors{Monson et al.}

\graphicspath{{./}{}}

\begin{document}

\title{Flare Induced Photospheric Velocity Diagnostics}

\correspondingauthor{Aaron Monson}
\email{amonson01@qub.ac.uk}

\author[0000-0002-3305-748X]{Aaron J. Monson}

\author{Mihalis Mathioudakis}
\affiliation{Astrophysics Research Centre, School of Mathematics and Physics,\\
Queen’s University Belfast, BT7 1NN,  \\
 Northern Ireland, UK}
 
 \author{Aaron Reid}
\affiliation{Astrophysics Research Centre, School of Mathematics and Physics,\\
Queen’s University Belfast, BT7 1NN,  \\
 Northern Ireland, UK}
 
\author{Ryan Milligan}
\affiliation{Astrophysics Research Centre, School of Mathematics and Physics,\\
Queen’s University Belfast, BT7 1NN,  \\
 Northern Ireland, UK}

\author{David Kuridze}
\affiliation{Department of Physics, Aberystwyth University, \\
Ceredigion, SY23 3BZ, UK}
\affiliation{Abastumani Astrophysical Observatory, Mount Kanobili, 0301 Abastumani, Georgia}

 \begin{abstract}

We present radiative hydrodynamic simulations of solar flares generated by the RADYN and RH codes to study the perturbations induced in photospheric Fe I lines by electron beam heating.  We investigate how variations in the beam parameters result in discernible differences in the induced photospheric velocities.  Line synthesis revealed a significant chromospheric contribution to the line profiles resulting in an apparent red asymmetry by as much as 40 m s$^{-1}$ close to the time of maximum beam heating which was not reflective of the upflow velocities that arose from the radiative hydrodynamic simulations at those times. The apparent redshift to the overall line profile was produced by significant chromospheric emission that was blueshifted by as much as 400 m s$^{-1}$ and fills in the blue side of the near stationary photospheric absorption profile. The velocity information that can be retrieved from photospheric line profiles during flares must therefore be treated with care to mitigate the effects of higher parts of the atmosphere providing an erroneous velocity signal.
 \end{abstract}


\section{Introduction} \label{sec:intro}

Solar flares  constitute  the  most  energetic  events  in  the  solar  system,  releasing as much as 10$^{32}$ erg of energy over tens of minutes \citep{Fletcher11}.  The standard flare model postulates that free magnetic energy, released by magnetic reconnection, accelerates particles in the newly realigned field lines, heating the plasma and conveying this energy to the lower solar atmosphere. This drives the acceleration of electrons to relativistic speeds creating a non-thermal beam that carries energy to the lower atmosphere \citep{Brown71,Emslie78,Holman11}. Observational evidence for this process is the emission of hard X-rays (HXR) from the footpoints of loops by the Bremsstrahlung process as the electrons collide with the dense lower atmosphere \citep{Hudson92,Neidig93}. The properties of the accelerated electrons can be inferred from the emitted HXR spectra \citep{Petrosian10}, provided by satellites such as RHESSI \citep{Lin02}, and can guide the creation of radiative-hydrodynamic models that provide insight into the response of the lower solar atmosphere to the heating.  Shocks and bulk plasma motions on the order of hundreds of km s$^{-1}$ are also generated as the chromosphere expands in a process known as chromospheric evaporation \citep{Milligan09}. While the majority of the flare energy is deposited into the upper chromosphere, leading to intensity enhancements and Doppler shifts of various lines, a significant amount can permeate to greater atmospheric depths \citep{Kosovichev98,Kosovichev07,Fletcher11}. X-ray and Extreme Ultraviolet (XEUV) backwarming is often assumed to be the dominant heating process in the photosphere and lower chromosphere as radiation emitted at higher geometric heights is reabsorbed in the denser medium \citep{Allred05,Lindsey08}. Radiative backwarming is shown to be largely dominated by the Balmer continuum and can drive photospheric heating \citep{Heinzel14,Kleint16,Kowalski17}. Proposed alternative heating methods include MeV protons that are accelerated alongside the electrons \citep{Neidig93,Prochazka18} or Alfvén waves \citep{Kerr16}. 

There have been a limited number of investigations into the effects of flare-induced  photospheric line-of-sight (LOS) velocities using radiative-hydrodynamic (RHD) simulations and the creation of the corresponding synthetic line profiles. The Fe I 617.3 nm line has been the focus of recent investigations given its deep formation height of around 200 km above the photospheric floor \cite[ $\tau_{500nm}=1$, ][]{Norton06,Hong18,Svanda18} and relevance to Solar Dynamics Observatory's Helioseismic and Magnetic Imager  (SDO/HMI) observations \citep{Pesnell12,Scherrer12}. \cite{Sharykin17} found that the X-class flare of 2012 October 23 intensity enhancements in the HMI filtergrams were co-spatial with the emission of HXRs but the maxima were delayed by up to 4 seconds relative to the peak of the HXR flux, consistent with predictions from radiative hydrodynamic simulations. They noted a weak downflow of around 0.5 km s$^{-1}$ from the Doppler shift of the 617.3 nm line during the period of beam heating. They also concluded that electron beams below the $10^{12}-10^{13}$ erg s$^{-1}$ cm$^{-2}$ energy range do not carry enough momentum to drive helioseismic waves.  \cite{Heinzel17} found that variations in the HMI continuum were dominated by hydrogen Paschen recombination continuum from the chromosphere, with a small contribution from free-free emission. \citet{Hong18} also compared the response of the 617.3 nm line for a quiet sun and a penumbra initial atmosphere to injected electron beams. They concluded that the cooler penumbra  exhibited greater heating of the lower atmosphere through a combination of backwarming in the photosphere and direct beam heating of the lower chromosphere \citep{Carlsson95}. This heating resulted in an increase in the line  contribution function  in the lower chromosphere. This created a significant brightening in the line core and showed blue asymmetries as a result of upflows. This was particularly pronounced for the beam with higher values of the low energy cutoff (E$_{c}$). \cite{Sadykov20} investigated how variations in the magnetic field strength affected the parameters derived from the HMI spectral line profiles. They found that the simulated observation of the LOS velocity deviated significantly from the actual velocities due to the time-dependent observing sequence of individual filtergrams. They also showed that the extent of the induced LOS velocity most strongly correlated with the flux of ~50 keV electrons

In this work we used a range of electron beams to explore how the beam parameters affect the velocity profiles of photospheric lines. We use radiative-hydrodynamic (RHD) modeling techniques to create synthetic line profiles for the Fe \RomanNumeralCaps{1} 617.3 nm line, along with the Fe \RomanNumeralCaps{1} 630.1 nm and 630.2 nm lines that are formed at similar photospheric depths. Section \ref{sec:RHD} describes the utilized RHD modeling codes. Section \ref{sec:EBPS} presents the key outcomes of a parameter study that was based on  various electron beams and their induced photospheric LOS velocities. Section \ref{sec:Line_Profiles} analyzes the changes in the deep forming Fe \RomanNumeralCaps{1} lines as a result of the flare. In Section \ref{sec:Discussion} we discuss the findings of the previous sections and presents our concluding remarks. 

\section{Radiative-Hydrodynamic Modeling}
\label{sec:RHD}

Due to the computational requirements of flare modelling the process is often split into two parts. Firstly, a full hydrodynamic response with radiative transfer is calculated to model the solar atmosphere's response to the energy deposition. The radiative transfer at this stage is limited only to the most important species that contribute to the energy balance via radiation losses. Radiative transfer and line synthesis can then be calculated from the first stage model to synthesize additional spectral lines of interest.

The RHD RADYN code of \cite{Carlsson92,Carlsson95,Carlsson97} was initially created to model the solar atmosphere's response to an energy input in the form of acoustic waves in the chromosphere, with modifications by \cite{Abbett99} to allow heating by nonthermal electrons. RADYN was further developed by \cite{Allred05, Allred15} to add a more accurate angle dependent Fokker-Planck solution to the beam particle diffusion and include the effects of X-ray and extreme ultraviolet (XEUV) backwarming. Importantly, RADYN employs the adaptive grid of \cite{Dorfi87}, that weights the concentration of gridpoints depending on the importance of resolving gradients of certain model parameters. In flare simulations the weighting is primarily given to temperature and velocity, to resolve the transition region and the formation of strong shocks formed in the loops, as well as the atomic level populations and NLTE (i.e. departures from Local Thermodynamic Equilibrium) densities for the radiative transfer calculations.  RADYN calculates the atomic level populations for a six-level hydrogen atom with continuum, a nine-level with helium atom with continuum, and a six-level Ca \RomanNumeralCaps{2} ion with continuum. This allows the calculation of various transitions that are important for the chromospheric energy balance \citep{Allred15}. For a full description of the current capabilities of RADYN the reader should refer to \citet{Allred15}, and references within, along with the improvements made to the Fokker-Planck solution \citep{Allred20}.
Due to the computational challenges imposed by calculating the full hydrodynamic solution with NLTE effects that are necessary for flare models, codes such as RADYN are typically limited to 1D. This approximation is  representative of a single loop to which the accelerated electrons are magnetically confined. While this is not the full treatment of an evidently 3D phenomenon, it is possible to create a multi-thread of 1D models that attempts to more accurately replicate the loop arcade across the footpoint \citep{Rubio16}. More recent efforts of simplified 3D cases of flare RHD models such as \citet{Kerr20} successfully overcome the 1D computational limit but are restricted to optically thin radiation. 

To synthesize the Fe \RomanNumeralCaps{1} lines we utilized the RH code of \cite{Uitenbroek01} which solves the equations of radiative transfer under statistical equilibrium. Snapshots from the RADYN outputs act as inputs for RH in order to obtain more complete radiative transfer solutions for a given atmosphere. Additional atomic species to those calculated in RADYN can also be included, with abundances relative to hydrogen, provided sufficient atomic data is supplied. As RH is a time-independent code, the history of the atmosphere is omitted for any given simulation, thereby losing the non-equilibrium (NE) atomic level populations that are important for a full radiative transfer solution of key chromospheric species. RH instead calculates its atomic level populations using statistical equilibrium (SE), utilizing the NE electron densities from RADYN. RH also allows for the line synthesis calculations to be conducted using Partial Frequency Redistribution (PRD) which is important for the Mg \RomanNumeralCaps{2} lines \citep{Kerr19PRD} but this was not utilized in this study.

%While there is a small difference in the intensities calculated using SE or NE, the gain of the PRD treatment from RH is more advantageous. The relative importance of these additional effects were investigated by \citet{Kerr19NEQ} for the Mg \RomanNumeralCaps{2} h \& k lines. They showed that while SE is suitable for most stages of the flare with the  Complete Frequency Distribution (CRD) vs PRD effects showed a 200\%--1000\% difference in intensities. While the SE verses NE effects showing a lesser difference but are still an important consideration for the initial phases of the flare. 

The combination of these two codes provides a powerful toolset for the analysis of flares and allows us to gain a greater understanding of the underlying hydrodynamic processes that create the observed features \citep{Rubio16,Kuridze16}.

\section{Electron Beam Parameter Study} \label{sec:EBPS}
The primary aim of this work is to investigate the velocities induced in photospheric line profiles during flares and the effect of different electron beam parameters on the induced photopsheric velocity profiles. Our parameter study utilized the 72 models of the F-CHROMA archive (\url{https://star.pst.qub.ac.uk/wiki/doku.php/public/solarmodels/start})  with varying values of the beam's total energy flux (F$_{tot}$), spectral index ($\delta$), and low energy cutoff (E$_c$), which are summarized in Table~\ref{tab:FCHROMA}. The F-CHROMA models utilize a pre-flare VAL-C atmosphere \citep{Vernazza81}, which is stationary, so all observed velocities are as a direct result of the deposited energy. The injected energy has a triangular temporal profile over the first 20s, followed by 30s of atmospheric relaxation.  For the purposes of this work we consider the temperature minimum region (TMR; 534~km above  $\tau_{500nm}=1)$  to be the upper boundary of the photosphere in the pre-flare atmosphere. This allows us to effectively represent the photosphere’s compaction according to the movement of the TMR downwards throughout the evolution of the atmosphere.

\begin{table}[h]
    \centering
    \begin{tabular}{|c|c|}
      \hline
      {\textbf{Beam Parameters}} & {\textbf{ Possible Values}} \\
      \hline
      \textbf{Total Flux (erg cm$^{-2}$)}   & {$3\times10^{10}$  \;  $1\times10^{11}$  \;  $3\times10^{11}$}  \\
      \hline
      \textbf{Low Energy Cutoff (keV)}                 & {10 \; \; 15 \; \; 20 \; \; 25}  \\
      \hline
      \textbf{Spectral Index}                     & {3 \; \; 4 \; \; 5 \; \; 6 \; \; 7 \; \; 8} \\
      \hline
    \end{tabular}
    
    \caption{ The three variable electron beam parameters included in the F-CHROMA model archive and their corresponding values. Different combinations of these values allow for 72 individual beam heating scenarios.}
   
   \label{tab:FCHROMA}
\end{table}

\pagebreak 

\subsection{Photospheric Velocities} \label{radynvelocity}

\begin{figure}[hb]
    \centering
    \includegraphics[width=0.95\linewidth]{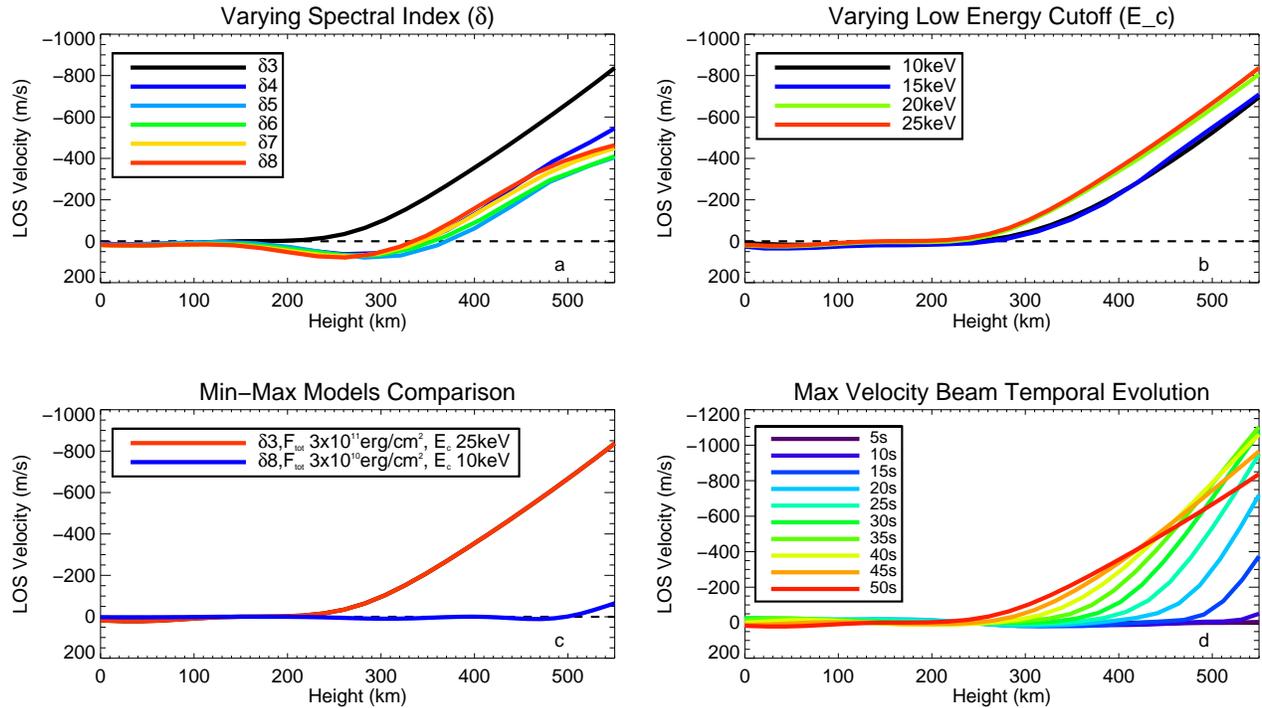}
    \caption{LOS velocities as a function of atmospheric height in the photosphere at the end of the RADYN simulations (50s) for the following beam parameters: a) Low energy cutoff E$_c$=25keV, $F_{tot}=3 \times 10^{11} $erg$ \ $cm$^{-2}$ and a varying spectral index ($\delta$). b) $\delta$=3, $F_{tot}=3 \times 10^{11} $erg$ \ $cm$^{-2}$ beam and a varying E$_c$. c) The models that show the highest  and lowest induced velocities. d) The temporal evolution of velocity for the beam that shows the highest upflow ($\delta=3$, E$_c=25$ keV, F$_{tot}=3 \! \times \! 10^{11} \ $erg$ \ $cm$^{-2}$) }
    \label{fig:radyn_velocity_plot}
\end{figure}

The general trend for all beam types was an induced upflow (negative velocities in Figure \ref{fig:radyn_velocity_plot}) in the upper photosphere as a result of heating. This is shown in Figures \ref{fig:radyn_velocity_plot}a and \ref{fig:radyn_velocity_plot}b, where the LOS velocity profile at the end of the models (50s) is presented as a function of the beam parameters. Variations in the spectral index (Figure \ref{fig:radyn_velocity_plot}a; F$_{tot}=3 \times \! 10^{11} \ $erg$ \ $cm$^{-2}$, E$_{c}$ = 25keV) showed that a majority of beams ($\delta$ = 4--8) have very similar  upflow velocities in the upper regions ($< -$600~m~s$^{-1}$). This upflow was balanced by a slower downflow in order to conserve momentum, originating at $\sim$~500~km during the periods of beam heating to a maximum downflow of 200 m s$^{-1}$. This downflow propagates lower into the atmosphere where the greater densities slow it to $\sim$ 100 m s$^{-1}$, and the overall photospheric velocity profile seen in Figure \ref{fig:radyn_velocity_plot}a at the end of model time (50s). The hardest beam type of $\delta=3$ (black line, Figure \ref{fig:radyn_velocity_plot}a) however showed a purely an upflow in the upper photosphere, with much greater velocities ($<$~1~km s$^{-1}$) at the upper boundary compared to the softer beams. This unique velocity profiles arises due to the greater flux of high-energy electrons in the $\delta$~= 3 distribution (discussed further in Section \ref{radynenergy}). 

Only the highest energy electrons within the beam can retain their energy deep into the atmosphere to heat the photosphere \citep{Fletcher07, Kennedy15}. Momentum is still conserved lower in the atmosphere for this  $\delta$ = 3 model, but the downflowing region originated at a greater depth with a higher density ($1 \! \times \! 10^{-7} \ $g$ \ $cm$^{-3}$ at 180 km compared to $3 \! \times 10^{-8} \  $g$ \ $cm$^{-3}$ for the other beam types' downflow formed higher) resulting in the  smaller downflow velocity.  Variations in E$_c$ further increase the upper boundary for the flux of the most energetic electrons, providing the staggered increasing depth for the beginning of the upflows for all beam types with increasing values for E$_c$. Additionally, the observed split of lesser upflows for the E$_c =$ 10 and 15 keV values in Figure \ref{fig:radyn_velocity_plot}b (F$_{tot}=3 \times \! 10^{11} \ $erg$ \ $cm$^{-2}$, $\delta$=3) was the result of a downwards propagating spike in density formed at $\sim$1200 km due to chromospheric condensation \citep[see also][]{Kowalski18}. This effectively creates a dense region of material that limits the penetration depths of high energy electrons and the subsequent heating, resulting in a reduced upflow velocity in the models that exhibit these condensations. Variations in the total flux of electrons within the beam act to scale the extent of the features denoted by varying the other two beam parameters, with  more energetic events resulting in  greater upflows (and corresponding downflow in relevant cases) velocity. This provides a huge range of potential induced velocities in the photosphere, with  the most extreme cases shown in Figure \ref{fig:radyn_velocity_plot}c. The temporal evolution of the model that showed the greatest upflow velocity ( $\delta$ = 3, E$_c$ = 25keV and F$_{tot} =3  \times \! 10^{11} \ $erg$ \ $cm$^{-2}$) is shown in Figure \ref{fig:radyn_velocity_plot}d. An accelerating upflow is induced for the initial 20s of the model when the energy is dissipated before reaching a maximum of $\sim$ 1 km s$^{-1}$ at 40s. Beyond this time the upflow velocity was seen to decrease as the lower atmosphere relaxed, with no clear evidence of the downflow that was observed for the other spectral indices.

\subsection{Energy Transport Mechanisms} \label{radynenergy}

\begin{figure}[h]
    \gridline{\fig{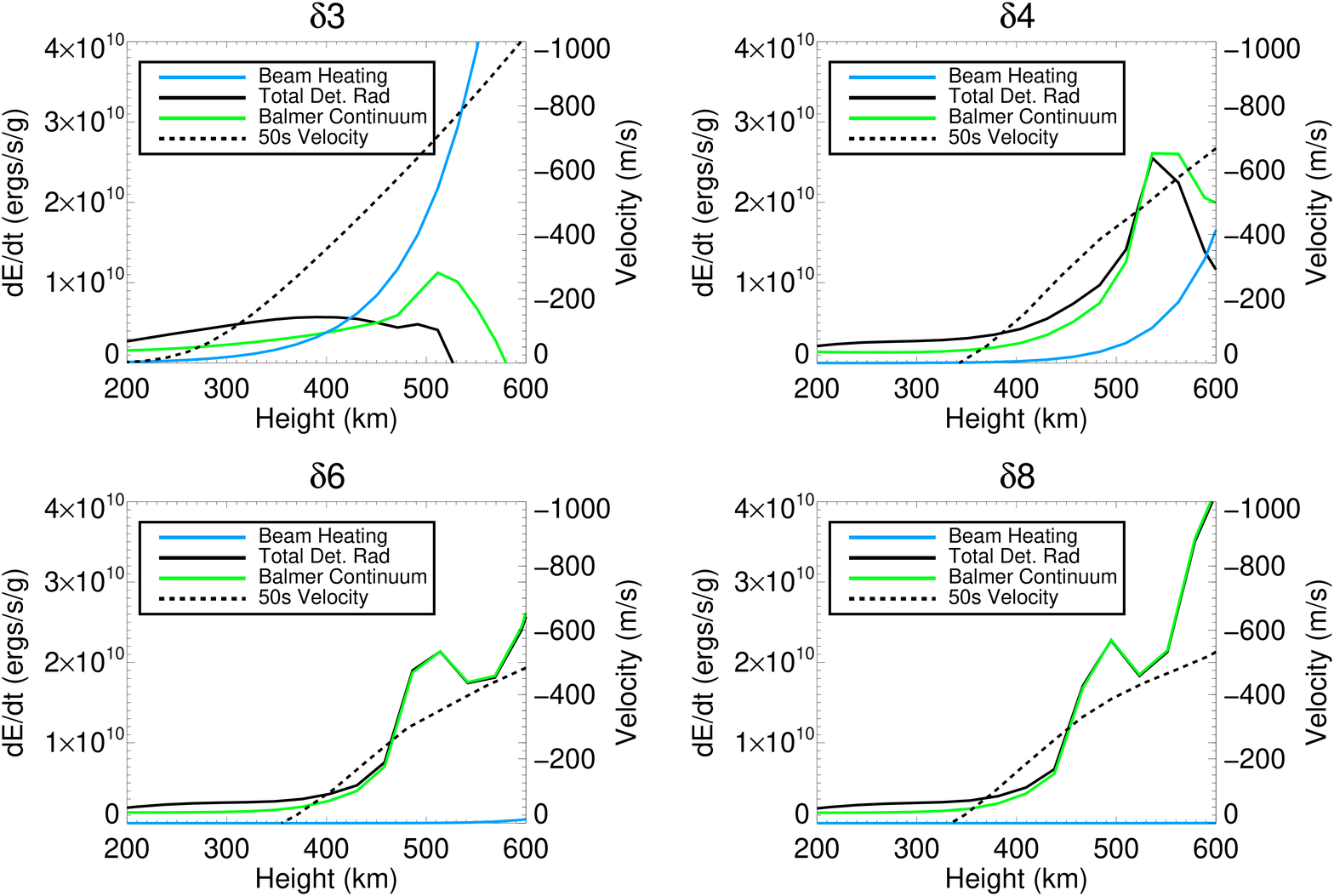}{0.51\linewidth}{(a)} \fig{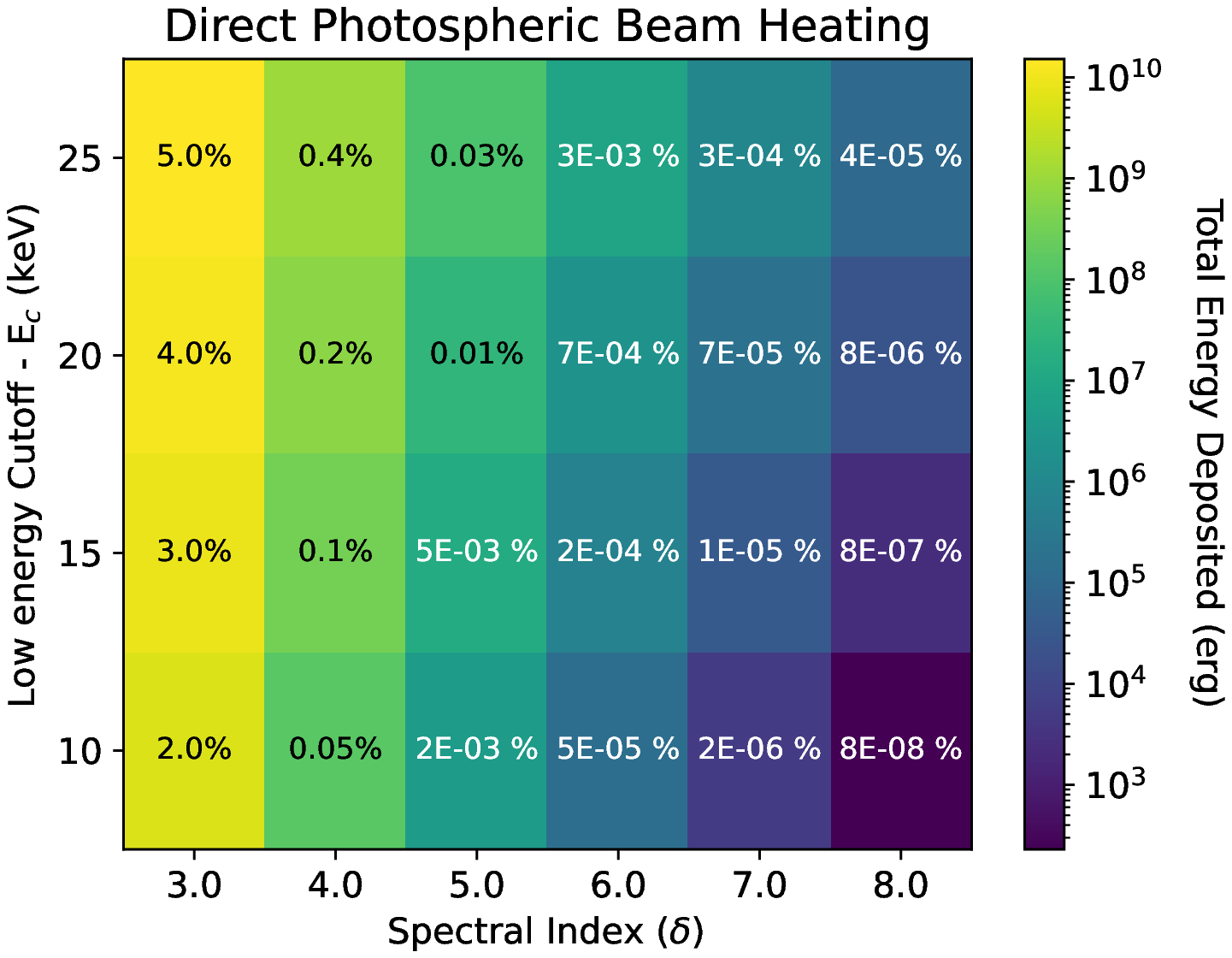}{0.48\linewidth}{(b)}}\centering
    
    \caption{Energy deposition in the photosphere as a result of electron beam heating. a) 
    The dominant heating terms at 10s (beam maximum) in the photosphere for various spectral index values of a E$_c$= 25 keV, $3 \times 10^{11} \ $erg$ \ $cm$^{-2}$ electron beam. The beam heating (blue line) is compared to the energy deposited by backwarming (black solid line), the major contributor of which is  Balmer continuum absorption (green line). The end of simulation LOS velocity for each case is also shown (black dashed line) to show the long term variations as a result of different heating profiles. b) The total energy deposited in the photosphere by direct beam heating for a $3 \times 10^{11} \ $erg$ \ $cm$^{-2}$ model for various values of $\delta$ and E$_c$. Percentages of the beam's total energy are overplotted for each cell.}
    \label{fig:energy_grid}
\end{figure}

Given the variations in the LOS velocities induced by different electron beams, it is necessary to examine the transport of energy to photospheric heights. In Figure \ref{fig:energy_grid} we show the dominant energy gains in the photosphere for a F$_{tot}=3 \times \! 10^{11} \ $erg$ \ $cm$^{-2}$ model with different $\delta$ and E$_c$, and how the variation in these parameters affects the direct beam heating of the photosphere. For the majority of cases, backwarming was found to dominate the energy gain and heating in the lower chromosphere/upper photosphere. The solid black line in Figure \ref{fig:energy_grid}a shows heating from absorption of all radiation detailed in the RADYN models (H, He, and Ca \RomanNumeralCaps{2} transitions and continua). A sufficiently hard ($\delta$=3 or 4) and energetic (F$_{tot}>1 \times 10^{11} \ $erg$ \ $cm$^{-2}$) beam can deposit energy deep into the atmosphere (blue line) and drive discernible differences in the LOS velocity profiles. This is further supported by the depth where the upflow originates in the $\delta$=3 beam type, which coincides with the maximum depth of where the beam heating permeates ($\sim$250 km, black dashed line for the upflow 50s velocity and blue colored beam heating profile on top left panel Figure \ref{fig:energy_grid}a). While the beam heating is not the dominant heating mechanism for the $\delta$=4 case, the direct beam heating still significantly augments the backwarming absorption to result in  the slightly higher upflow shown in Figure \ref{fig:radyn_velocity_plot}a for the blue line.  For beams softer than $\delta$=4, the alike velocity profiles arise as a result of the similar extent of XEUV backwarming that predominately drives the heating in the upper photosphere (400--534 km; the previously defined quiet sun photosphere upper boundary height).  This is caused mainly by the absorption of Balmer continuum radiation shown by the similar profiles of the green lines for the $\delta$=4, 6, and 8 beams in Figure \ref{fig:energy_grid}a that dominates much of the total backwarming \citep{Kowalski17}.

The $\delta$=3 case however provides a sufficiently high flux of high energy electrons to dominate over the XEUV backwarming heating profile seen in softer beams, resulting in the strong purely upflowing velocity profile exhibited in Figure \ref{fig:radyn_velocity_plot}a. Figure \ref{fig:energy_grid}b shows how the proportion of a beam's total heating that reaches photospheric heights varies with $\delta$ and $E_c$ for an energy of F$_{tot}=3 \! \times \! 10^{11} \ $erg$ \ $cm$^{-2}$. It shows that a maximum of only a few percent of the beam's energy can make it to the photosphere to cause substantial direct heating. The $\delta$ = 3 and 4 cases in Figure \ref{fig:energy_grid}a where beam heating is a significant heating factor compared to the backwarming effect.

\section{Synthesized Line Profiles}
\label{sec:Line_Profiles}

\begin{figure}[h]
    \centering
    \includegraphics[width=\linewidth]{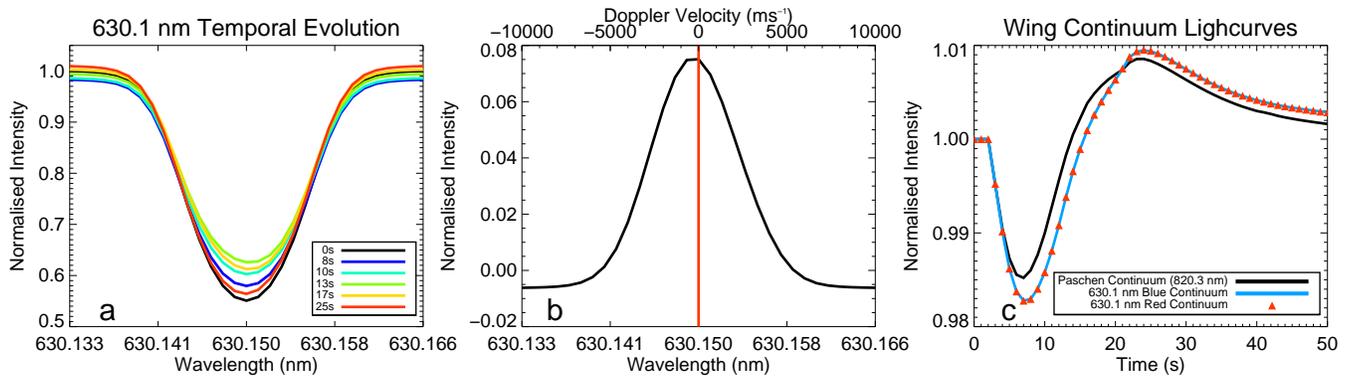}
    \caption{a) The temporal evolution of the RH synthesized 630.1 nm line normalized to the pre-flare continuum levels. b) The net emission profile derived by subtracting the pre-flare profile from the one at 13s that shows the maximum core brightness. The excess emission shows a blue asymmetry corresponding to a Doppler velocity of -500m s$^{-1}$ with negative velocities indicating an upflow. c) The temporal evolution of the normalized red and blue continua on either side of 630.1 nm line compared to a point in the Paschen continuum at 820.3nm.}
    \label{fig:RH_Intensity_plot}
\end{figure}

In this section we focus on the RADYN model that exhibited the greatest induced velocity profiles and direct beam heating in the photosphere ($\delta=3$, E$_c=25$ keV, F$_{tot}=3 \! \times \! 10^{11} \ $erg$ \ $cm$^{-2}$). This model was exported to RH at one second cadence for the full 50s of the simulations. The NLTE calculations were performed for the Fe \RomanNumeralCaps{1} lines: 617.3 nm, 630.1 nm, and 630.2 nm to investigate the Doppler shifts that can be retrieved from these lines given the velocity structure induced by the electron beam as discussed in Section \ref{radynvelocity}. For the purpose of this work we shall focus on the analysis of the 630.1 nm line that showed the greatest extent of the effects that we will discuss. We emphasize that all three transitions show similar trends. The temporal evolution of the 630.1 nm line profile, normalized to the pre-flare continuum level at 630.133 nm, is shown in Figure \ref{fig:RH_Intensity_plot}a. The line core showed a maximum brightening of $\sim$10\% during the period of beam heating. The maximum core intensity occurred at 13s, shortly after the maximum beam heating at 10s and is consistent with the several second delay seen by \citet{Sharykin17}. Following the end of beam heating the line core intensity rapidly returned to its pre-flare level (red line in Figure \ref{fig:RH_Intensity_plot}a shows a near pre-flare intensity shortly after the end of beam heating).

A Gaussian fit of the line profile at each timestep revealed a redshift during the initial 20s of the model,  followed by a  blueshift as the line profile returned to a near pre-flare level. However, this is in contrast to the velocities from the RADYN model atmosphere in two regards: firstly, as discussed in Section \ref{radynvelocity} and shown in Figure \ref{fig:radyn_velocity_plot}d, this particular RADYN model exhibited no clear downflows in the photosphere during the initial stages of the beam heating that would induce a redshift in the model and discussed in \citet{Sharykin17} for the 617.3 nm line. This led the previous work of \citet{Hong18}'s FQa model to denote this as a "fake" red asymmetry. The redshift seen during the initial 20s of the beam  appears to be the result of a blue-sided asymmetry that becomes apparent when we subtract the pre-flare intensity profile from the flare profile, an example of which is shown in Figure \ref{fig:RH_Intensity_plot}b for the time of maximum brightness at 13s. The asymmetric filling in of the blue side of the profile, created by the upflow of plasma seen in the RADYN models resulted in an apparent redshift that could be mistakenly attributed as a downflow created by the beam. However, the source of this filling in cannot be photospheric in origin. At this time in the RADYN model (13s) no point in the photosphere shows a LOS velocity close to the $\sim$ $-$400 m s$^{-1}$ needed to create the excess shown in Figure \ref{fig:RH_Intensity_plot}b. Secondly, despite the upper photosphere's upflow velocity generally increasing over time, the blueshift velocities exhibited by the bisectors of the 630.1~nm line following the end of beam heating ($<$100m s$^{-1}$) were still greater than the upflow velocity in the lower photosphere. In the 0--300 km region, where 95\% of the line is formed in the quiet sun the velocity did not exceed 50 m s$^{-1}$ (Figure \ref{fig:radyn_velocity_plot}d). We did not see an deep photospheric upflow velocity of a magnitude that can create the blueshift seen in the line profile . 

To rule out the possibility that continuum enhancements could create the observed changes, the red and blue continuum intensities were compared to the  change in the Paschen continuum intensities throughout the 50s of the model. Lightcurves for the 630.1 nm red and blue wing continuum intensities, and a point in the Paschen continuum (820.3nm), are shown in Figure \ref{fig:RH_Intensity_plot}c with all values normalized to their pre-flare levels. The initial dip in the wings of the line shown in Figure \ref{fig:RH_Intensity_plot}a is due to a suppression of the intensity in the Paschen continuum \citep{Abbett99,Allred05}, closely following the lightcurve for the continuum in Figure \ref{fig:RH_Intensity_plot}c. Furthermore, Figure \ref{fig:RH_Intensity_plot}c showed that continuum variations cannot be the cause of the blue excess in the line core. While the line core intensity for the 630.1~nm line increases in the initial 15s of the energy injection (Figure \ref{fig:RH_Intensity_plot}a), the Paschen continuum shows a dip in intensity, an overall variance of $\sim 1\%$. This continuum intensity decrease during the period of core brightening, and the scale of this variance not being on the order of 10\% variance needed to explain the core enhancement means it could not be the source of the signal of the blue excess noted during these times. 

\subsection{Quantifying the Chromospheric Contribution to the Fe I Lines} \label{sec:Contribution}

\begin{figure}[b]
    \gridline{\fig{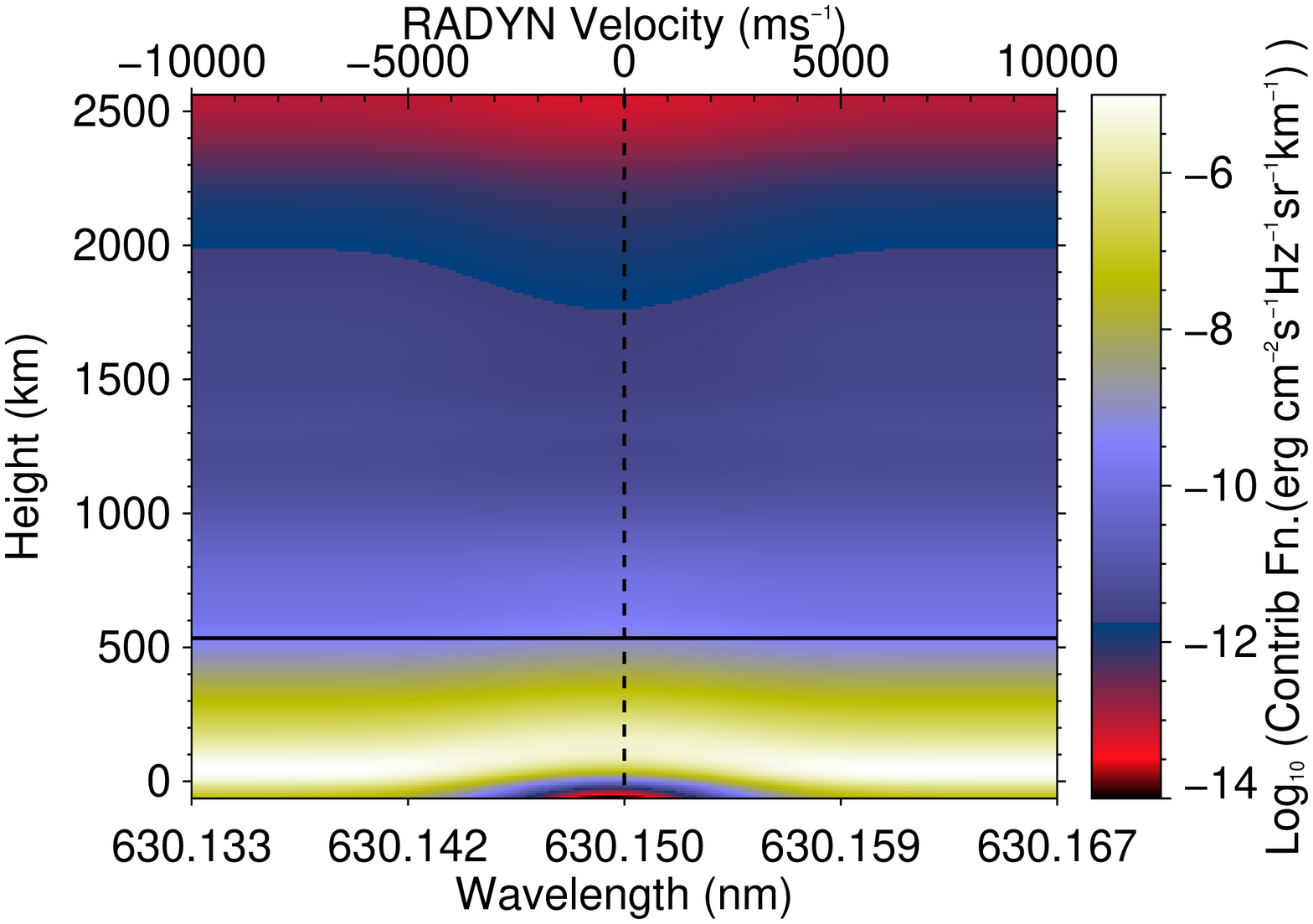}{0.48\textwidth}{(a)} 
                \fig{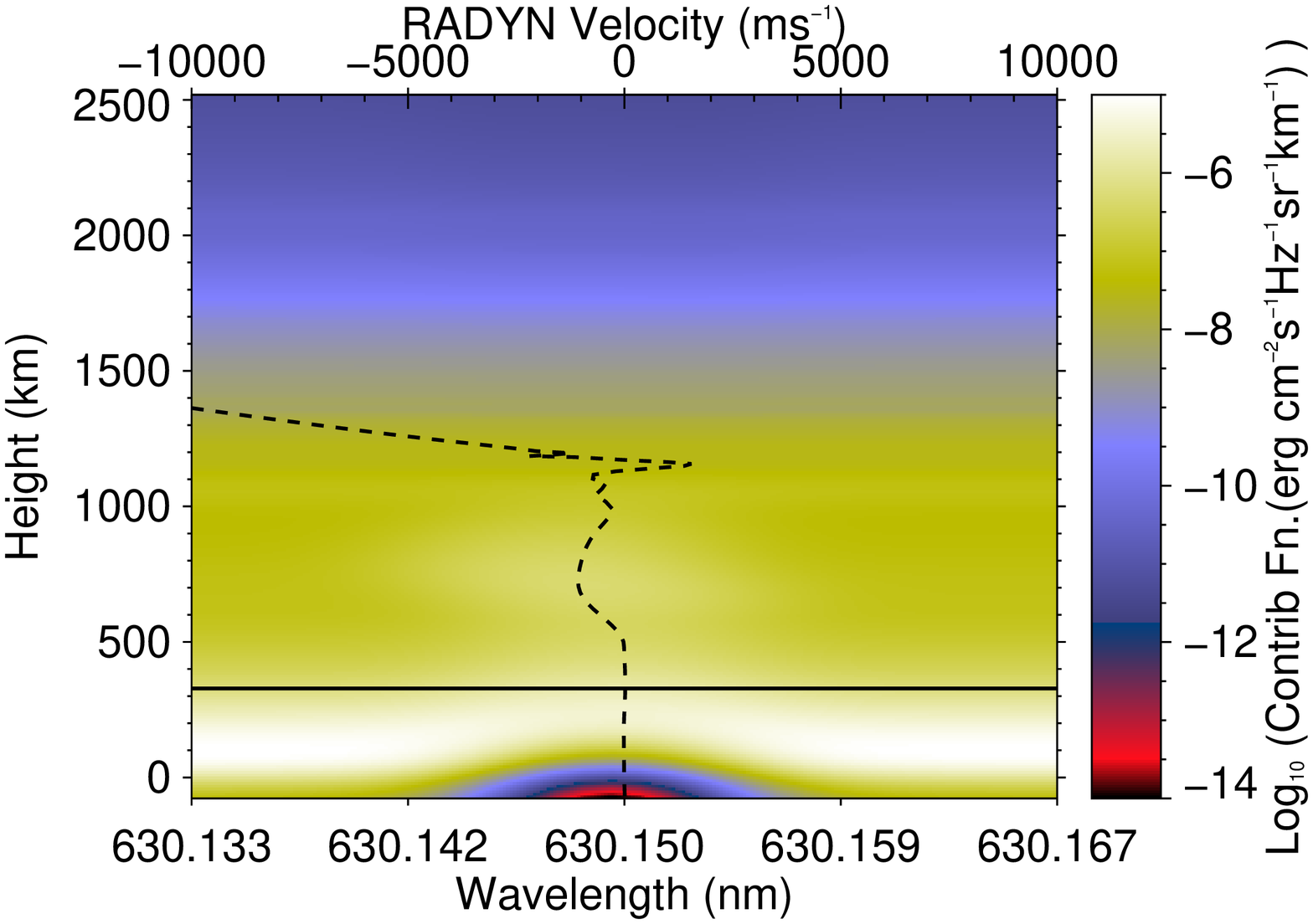}{0.48\textwidth}{(b)}}
    \caption{Contribution functions as a function of height for a) the pre-flare  and b) the 13s maximum core brightness time atmosphere for the Fe I 630.1 nm line profile. The corresponding velocity from RADYN is shown as a dashed line with negative values corresponding to upflows. The TMR (upper boundary for the photosphere) is marked as a horizontal solid black line.} 
    \label{fig:Contrib_panels}
\end{figure}

To resolve the counter-intuitive velocity signature illustrated by Figure \ref{fig:RH_Intensity_plot}, we investigated the contribution functions \citep{Carlsson95} described in Equation \ref{eq:contrib_function} to disentangle the velocity signal. 
\begin{equation}\label{eq:contrib_function}
\centering
 I_\nu = \int_{z_{\small0}}^{z_{\small1}} C_{\nu} \,dz =  \int_{z_{\small0}}^{z_{\small1}}S_{\nu} \, \tau_{\nu} \, e^{-\tau_{\nu}} \, \frac{\chi_{\nu}}{\tau_{\nu}} \,dz
\end{equation}

Equation 1 shows that the emergent intensity at a given frequency (I$_{\nu}$) can be calculated by integrating the contribution function (C$_{\nu}$) over the atmospheric height scale. The contribution function is given by the product of the source function (S$_{\nu}$), optical depth ($\tau_{\nu}$) and density of emitters ($\chi_{\nu}$) at a given frequency. This provides an effective tool to help visualize, on a height scale, where in the atmosphere a given transition is formed and the effect of the hydrodynamic properties of the material on an observed line profile \citep{Kuridze16,Kerr19PRD}. The contribution function for the 630.1~nm line is shown in Figure \ref{fig:Contrib_panels}a for the pre-flare atmosphere. The corresponding velocity profile from RADYN is given as a dashed line, which is stationary in the pre-flare state. The photospheric upper boundary, defined as the location of the temperature minimum (534~km), is marked as the horizontal solid black line. The contribution function shows that the Fe I 630.1~nm line is almost entirely photospheric in origin, with 95\% of its formation attributed to the 0--300 km region. The effect of the flare's energy deposition in the chromosphere is shown in Figure \ref{fig:Contrib_panels}b as a visible increase in the contribution function across the entire line profile. This is most notable in the 550--900 km range, highlighted by the increase in the upflow velocity (negative velocities) from RADYN in this region, but extends up to 1600 km,  above where the peak beam energy deposition occurs. At direct viewing angles (in our case $\mu$=0.887) this chromospheric surplus has a significant effect on the velocity information that can be retrieved.

\begin{figure}[b]
    \centering
    \includegraphics[width=\linewidth]{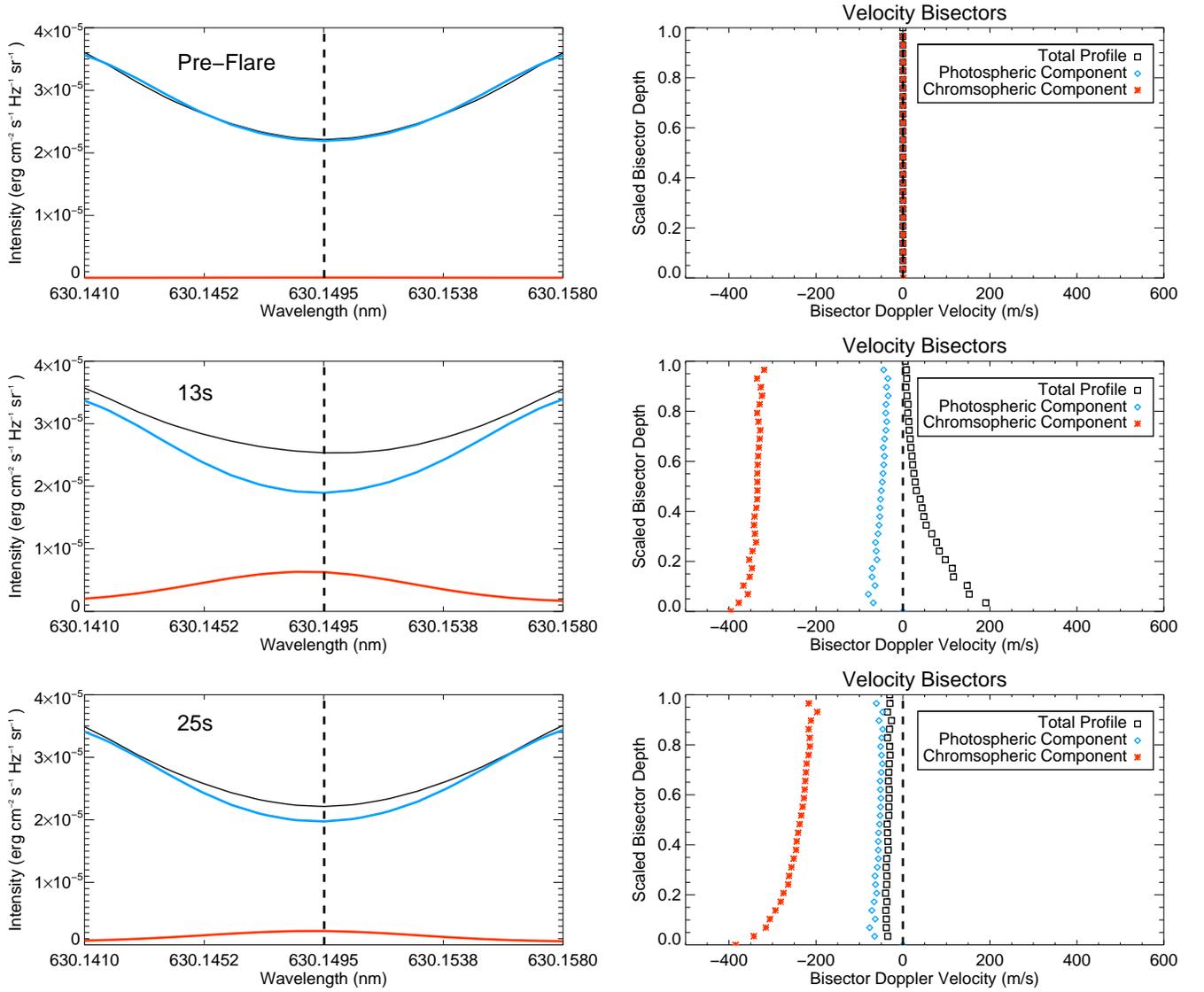}
    \caption{ Left Panels: Intensities calculated from the integrated line contribution functions. Right panels: The velocity bisectors for the total emergent intensity and its two components at the same times. Negative velocities correspond to blueshifts. \textit{Top Row:} The pre-flare atmosphere. \textit{Middle Row:} The time of maximum core brightness (13s). \textit{Bottom Row:} A time after beam heating has ceased (25s).}
    \label{fig:Profile_Analysis}
\end{figure}

 As the emergent intensity from a given region can be estimated by integrating the contribution function between two heights within the total range, we split the total emergent line profile into photospheric and chromospheric components. This was performed with a temporal resolution of 1 second to track the effect of the excess chromospheric  contribution on the core of the 630.1~nm line. Figure \ref{fig:Profile_Analysis} shows for three representative cases: the pre-flare state, the time of maximum core brightness (13s), and at a time after the beam heating has ended (25s). The corresponding velocity bisectors are also plotted (right panels) scaled between each component's maximum and minimum intensity. In the pre-flare atmosphere the total emergent profile (black) is entirely due to the photospheric component (blue line) with negligible chromospheric contribution (red), reflective of the contribution function seen in Figure \ref{fig:Contrib_panels}a. The velocity bisectors at this time reflect the initial stationary VAL-C atmosphere. However, during the period of beam heating we can see a significant chromospheric contribution to the total emergent intensity (middle row) in addition to the photospheric pre-flare profile. This chromospheric excess is the primary reason for the profile's increased brightness during these times while the photospheric component remains fairly constant. The velocity bisectors at this time show a slow ($< -100$ m s$^{-1}$) photospheric Doppler velocity (blue diamonds) and a faster ($< -400$ m s$^{-1}$) chromospheric  (red stars) blueshift of the two components. While the photospheric component's bisectors are reflective of the velocities in the upper photosphere at this time (Figure \ref{fig:radyn_velocity_plot}d), the faster moving chromospheric excess results in the uneven filling in to the blue side of the line core, creating an apparent redshift to the total emergent profile as can be seen by its velocity bisectors (black squares). The total emergent intensity profile does not reflect the blueshifted photospheric and chromospheric components that constitute it. Following the end of beam heating the chromospheric contribution diminishes in intensity and the total profile restored back to the photospheric component (25s panel). Although the chromospheric component still showed blue shifted bisectors of several hundred m s$^{-1}$, its diminished intensity contribution to the total line profile led to a marginal blueshift of $\sim -40$ m s$^{-1}$ similar to that shown for the photospheric component.

\begin{figure}[b]
    \centering
    \includegraphics[width=\linewidth]{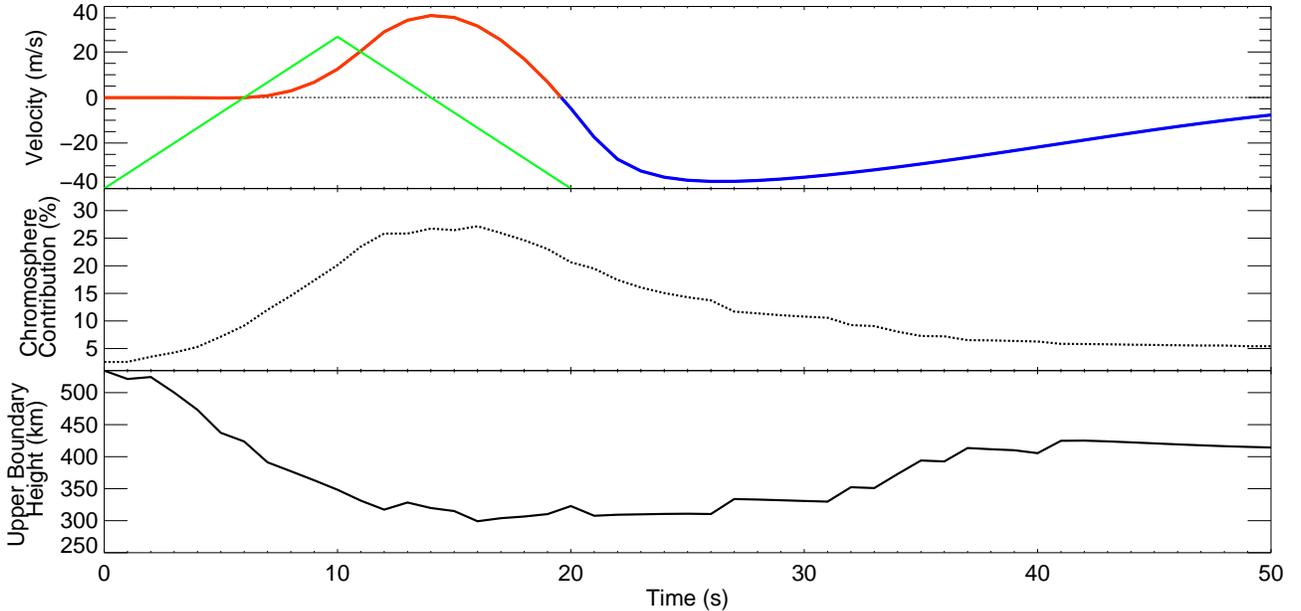}
    \caption{Breakdown of Doppler velocities derived from the Gaussian fits to the the 630.1 nm line. \textit{Top Panel}: The Doppler shift of the line core over the 50s of the simulation. Negative velocities corresponding to blueshifts. The temporal profile of the injected electron beam flux (green triangle) is provided for context. \textit{Middle Panel}: The contribution of the chromospheric region to the total emergent line core intensity carried out by integrating the line contribution function. \textit{Bottom panel}: The variation of the photospheric upper boundary defined by the location of the TMR.}
    \label{fig:6301_three_panel}
\end{figure}

The resultant effect of this chromospheric contribution during the period of beam heating is the masking of the true photospheric velocities at these times. It is seen however in the bottom row of Figure \ref{fig:Profile_Analysis} that after the end of beam heating it is possible to observe blueshifts of the 630.1~nm line that are in agreement with the RADYN model upflows discussed in Section \ref{sec:EBPS} within the same order of magnitude.
The temporal evolution of the 630.1~nm line core velocity obtained from the Gaussian core fit is shown in Figure \ref{fig:6301_three_panel} (top panel). A peak redshifted velocity of $\sim$40 m s$^{-1}$  coincided with the times of the maximum chromospheric contribution  to the total intensity profile (middle panel) in the 13--15s range, a few seconds after the time of the peak energy deposition. At its peak, the chromospheric component contributed up to 25\% of the line core intensity, and once the chromospheric contribution began to recede, a blueshifted profile emerged. The switch from the initial redshift to a blueshifted profile was due to the combined effects of i) a reduced chromospheric contribution to the line core following the end of beam heating and ii) the photosphere's upflow sufficiently evolving to greater velocities and greater depths as shown in Figure \ref{fig:radyn_velocity_plot}d.

While the variation in the height of the photosphere's upper boundary plays a role in what is deemed photospheric or chromospheric in origin throughout the 50s, affixing the photospheric upper boundary to the pre-flare height (534~km) results in a maximum of  10\% of the total emergent intensity being re-classified as remaining photospheric throughout the simulation. This still leaves up to 15\% of the total profile being contributed from the higher formation region of 550--1200 km. A small increase in the chromospheric contribution at the end of the models (Figure \ref{fig:6301_three_panel}, middle panel: 50s) compared to the pre-flare atmosphere, is attributed to the simulation ending before the atmosphere had fully returned to its pre-flare state.

We therefore see that for a narrow time range (20--30s) the blueshift exhibited in the 630.1~nm line is representative of the velocities in the photosphere, indicating an upflow on the order of 10’s m s$^{-1}$ (accurate for the mid-photosphere velocities in Figure \ref{fig:radyn_velocity_plot}d in this time range). Following this the contribution function returns to a near pre-flare state, with the line profile being formed almost entirely within the 0--300 km region. The exhibited blueshift in the line profile diminished during the late stages of the model (top panel in Figure \ref{fig:6302_three_panel}; 30--50s) as the line is once again formed at greater geometrical depths and below the upflows seen in the upper photosphere.

\section{Discussion and Conclusions}
\label{sec:Discussion}

 We used radiative-hydrodynamic simulations to investigate the viability of retrieving accurate flare-induced photospheric line-of-sight velocities from the Fe \RomanNumeralCaps{1} 630.1~nm line. Our study revealed that while the line is formed in the lower photosphere in the quiet sun, a significant chromospheric contribution emerges during the periods of electron beam heating leading to a filling-in effect that increases the intensity of the line core. The chromospheric component creates an apparent redshift to the total emergent line profile due to the asymmetric filling-in of the line core. The fast-moving blueshifted chromospheric component masked the true photospheric velocities, hence negating the use of this line in inferring photospheric velocities during the period of beam heating. Our analysis has been extended to include the Fe \RomanNumeralCaps{1} 617.3~nm and 630.2~nm lines, which also originate from the photosphere in the quiet sun, and show results similar to the Fe \RomanNumeralCaps{1} 630.1~nm (Appendix A).

We investigated how variations in the parameters of a beam of accelerated electrons resulted in discernibly different photospheric LOS velocities. Our study revealed that while the total energy of a given beam is important in determining the scale of the photospheric response, variations in the spectral index and low energy cutoff also contribute to discernibly different photospheric velocity profiles. A parameter study of the 72 models of the F-CHROMA archive has shown a general trend of upflows in the upper photosphere on the order of several hundred m s$^{-1}$ accompanied by a downflow at greater depths due to  conservation of  momentum. The smaller magnitude of these photospheric velocities, compared to the km s$^{-1}$ values encountered in the chromosphere, is anticipated given the denser medium and greater geometric length that the electrons travel through in the lower atmosphere. 

In the case of higher energy beams  (F$_{tot} = 3 \times 10^{11}$ erg cm$^{-2}$ and E$_c$ = 25~keV) we found that all but one of the spectral indices  showed a clear pattern of upflows accompanied by downflows at lower heights, with a  very similar grouping in the induced velocity profiles (Figure \ref{fig:radyn_velocity_plot}a).  Unique to this trend is the hardest beam with $\delta$=3 that shows a much greater upflows approaching 1 km s$^{-1}$ that originated deeper into the atmosphere than the models with higher spectral indices. We believe that the similar grouping of velocities for the $\delta$= 4--8 beams is caused by a common primary heating mechanism in the photosphere, which is backwarming. This is in agreement with the findings of \citet{Hong18} who concluded that below 500~km backwarming is generally the dominant energy heating process. This backwarming seems to be  dominated almost entirely by the Balmer continuum consistent with \citet{Kowalski17} (Figure \ref{fig:energy_grid}a; $\delta$=4, 6, and 8 panels). These almost identical heating profiles for differing values of the spectral index is what results in the grouping of the induced velocities shown in Figure \ref{fig:radyn_velocity_plot}a.

The unique velocity profile induced for the hardest beam type of $\delta$=3 is a result of direct beam heating dominating over backwarming. The deeper origin for the upflow in this case was also found to correspond to the greatest depth the beam is able to penetrate (shown in Figure \ref{fig:energy_grid}a). Increasing the low-energy cutoff further extends the maximum penetration depth due to the larger number of high energy electrons.  Furthermore, it is interesting to note that chromospheric features such as the chromospheric compressions induced in the models with lower values of E$_c$ can have a noticeable effect on the photospheric velocity profiles. These dense shocks created in the lower chromosphere, created an effective barrier that limits the extent of direct beam heating at greater depths. This is shown in Figure \ref{fig:radyn_velocity_plot}b as a grouping of velocity profiles of models that do and do not induce these compressions,resulting in differening maximum upflow velocities. The variations in E$_c$ become increasingly relevant at higher values of $\delta$, providing a variation of several orders of magnitude in the deposited energy for the softest beams (Figure \ref{fig:energy_grid}). This highlights that variations in the spectral index and low energy cutoff result in greater variations in the photospheric energy deposition than variations in the total energy of the electron beam. Excluding the hardest beam types ($\delta$=3 or 4) direct beam heating becomes insignificant compared to the XEUV backwarming effects. This would indicate that the flare classification (X/M/C...) is not always indicative of the photospheric response. For example, a beam with F$_{tot}=10^{10} \ $erg$ \ $cm$^{-2}$, $\delta=3$ could have a greater photospheric energy deposition than one with  F$_{tot}=10^{12} \ $erg$ \ $cm$^{-2}$, $\delta=5$. While the $\delta$=3 beam type is an extreme case we note that softer beams, such as those with a $\delta=4$, but with a sufficiently high Ec ($\sim 50-60$ keV) can contain a significant flux of high energy electrons to drive a similar extent of direct beam heating at photospheric heights as those driven by a harder beam.  

Following our investigation of the F-CHROMA grid, we conclude that the $\delta=3$, E$_c=25$ keV, F$_{tot}=3 \! \times \! 10^{11} \ $erg$ \ $cm$^{-2}$ model is the one most likely to create an observable change in spectral lines forming deep in the solar atmosphere.  This choice of beam parameters is in agreement with previous works (\citet{Hong18,Sadykov20}. We have also explored how variations in these parameters may affect the key factors which would cause observable changes in photospheric spectral lines; the induced LOS velocity and energy deposition mechanisms. We used this electron beam for the synthesis of photospheric spectral lines. 

We synthesized profiles for the photospheric lines of  Fe \RomanNumeralCaps{1} (617.3~nm, 630.1~nm, and 630.2~nm) at 1s cadence to investigate the velocity information that could be retrieved. We focused on the 630.1~nm line due to it exhibiting the largest magnitude of the discussed effects, with the 617.3~nm and 630.2~nm lines showing similar trends. During the period of beam heating all three lines exhibit a brightening in their line cores of $\sim$10\%, peaking several seconds after the maximum beam time (617.3~nm;15s, 630.1~nm;13s and 630.2~nm;14s). The staggered timing in the maximum brightening is a result of the 630.1~nm  contribution function peaking slightly higher in the atmosphere ($\sim$120~km) than the other two lines ($\sim$50~km). The response time for all three lines however is consistent with the delay between the HXR peak (representative of the peak beam heating time) and the photospheric disturbances described in \citet{Sharykin17}. Following the end of beam heating all three lines rapidly return to their pre-flare intensities, as shown in Figure \ref{fig:RH_Intensity_plot}a by the 25s mark the profile is near identical to the pre-flare profile. 

During the times of increased core brightening, a blue asymmetry was detected (Figure \ref{fig:RH_Intensity_plot}b) that was far in excess of the velocities in the photosphere. This blue excess caused an apparent redshift of the total line profile as the blue side of the emergent profile was filled in. This could be  easily misinterpreted as a downflow in the photosphere created by the momentum of the nonthermal electrons, as in \citet{Sharykin17}. Analysis of the line contribution functions revealed that the blue excess originated in the chromosphere during the period of most intense beam heating (Figure \ref{fig:Contrib_panels}, in agreement with \citet{Hong18}). 
The chromosphere contributed up to 25\% of the 630.1 nm line core intensity with the 617.3 nm and 630.2 nm chromospheric contributions at 15\% and 14\% respectively. The evolution of velocity for these two lines are included in the appendix. Following the end of beam heating there was still a sufficient blueshifted contribution from the upper photosphere that is reflective of the RADYN upflow in the region. The return of the contribution function to lower atmospheric heights provides the diminishing blueshifted signal in Figure \ref{fig:6301_three_panel}a. This limits the velocity diagnostic potential of these lines later in the simulation. While the largest photospheric velocities occurred at the late stages of the simulation, the three lines do not reflect this as their formation height is once again almost entirely in the 0--300km region.

We therefore conclude that any attempts to infer photospheric LOS velocities from the Doppler shifts of spectral lines that are formed in the photosphere must account for the chromospheric contribution to ensure the correct attribution of the inferred velocities. Even in the case of an electron beam causing a pure upflow in the photosphere, such as the model used in our line synthesis, accurate Doppler velocities are not retrievable. %How this chromospheric contribution varies and affects the observed Doppler shifting of deep forming spectral lines in other electron beam parameter combinations will be the focus of future work. 
Future work in this area could investigate the effects of particle beams on the photospheric spectral lines of active late-type stars. These objects generate very energetic flares with very hard X-ray spectra and total energies up to 10,000 more than their solar counterparts \citep{Osten10,Kowalski15}, providing an excellent potential for significant photospheric reactions.

\acknowledgments

AJM would like to thank Han Uitenbroek, Graham Kerr, Ivan  Mili\'{c} and Adam Kowalski for their discussion on RH contribution functions.  We would like to thank the anonymous referee for useful comments and suggestions. This research is based on the grid of flare models produced under funding from the European Community’s Seventh Framework Programme (FP7/2007-2013) under grant agreement no. 606862 (F-CHROMA), and from the Research Council of Norway through the Programme for Supercomputing. AJM acknowledges funding from the Science Technology Funding Council (STFC) Grant Code ST/T506369/1. MM and AR acknowledge support from the Science and Technology Facilities Council (STFC) under grant No. ST/P000304/1 \& ST/T00021X/1. RM would like to thank Science and Technologies Facilities Council (UK) for the award of an Ernest Rutherford Fellowship (ST/N004981/2). DK has received funding from the S\^{e}r Cymru II scheme, part-funded by the European Regional Development Fund through the Welsh Government. 

\appendix

\section{Velocities derived from the Fe I 617.3 nm and 630.2 nm lines}

\begin{figure}[h]
    \centering
    \includegraphics[width=0.98\linewidth]{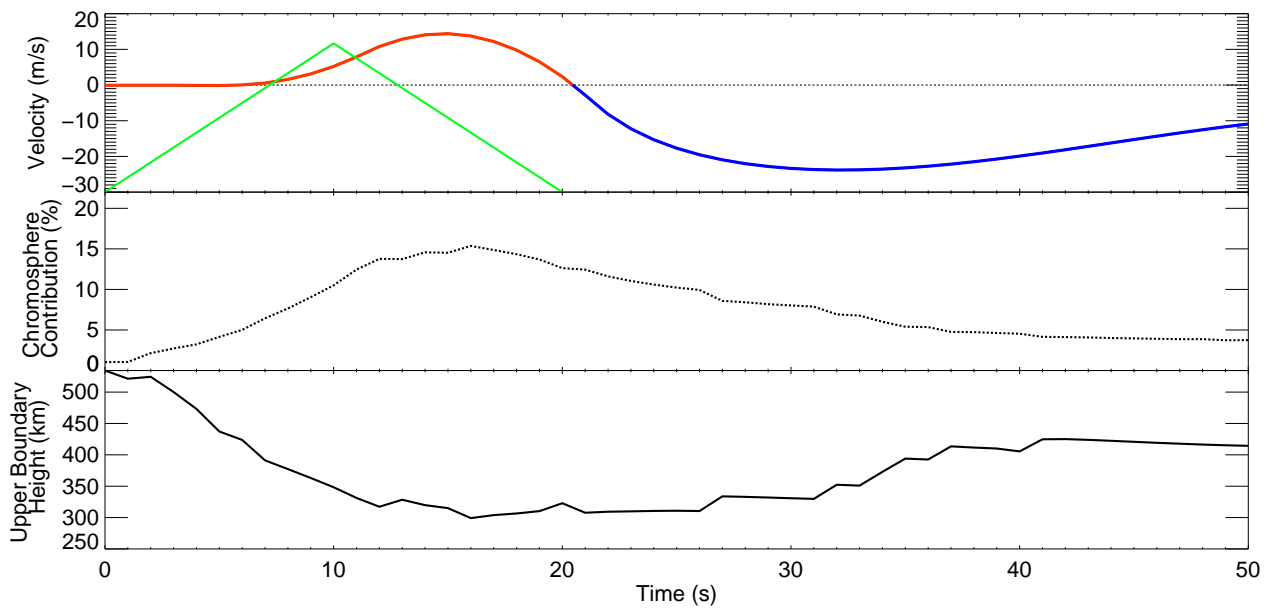}
    \caption{The LOS velocities of 617.3 nm during the 50s of the simulations. Description as in Figure \ref{fig:6301_three_panel}.}
    \label{fig:6173_three_panel}
\end{figure}

\begin{figure}[h]
    \centering
    \includegraphics[width=0.98\linewidth]{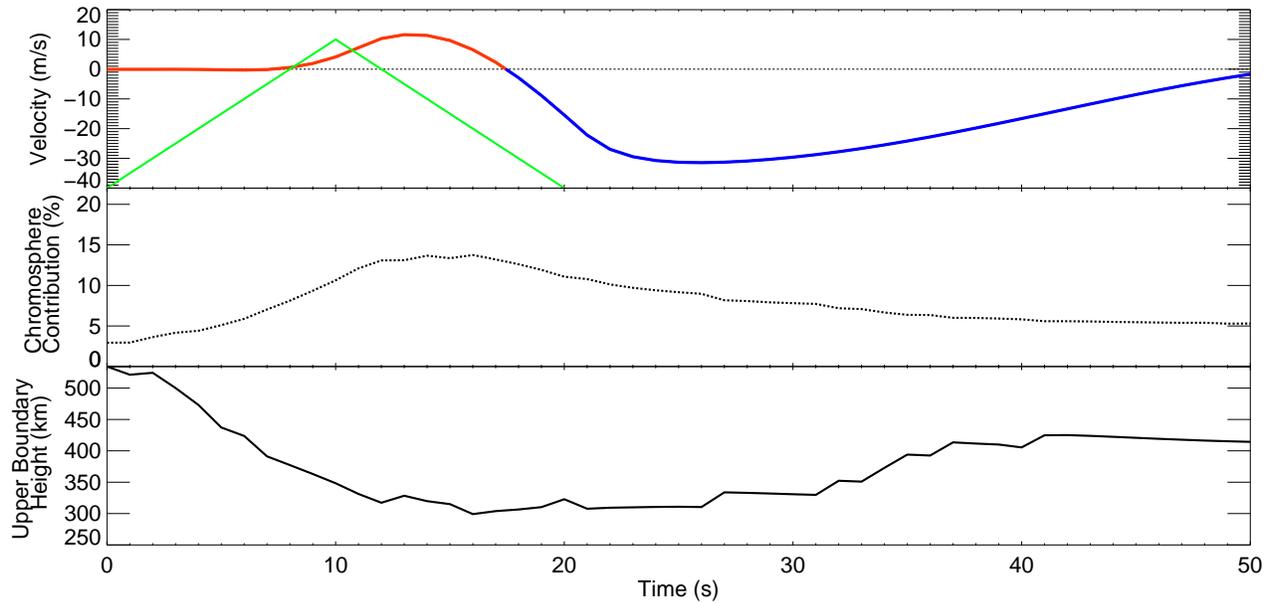}
    \caption{The LOS velocities of 630.2 nm during the 50s of the simulations. Description as in Figure \ref{fig:6301_three_panel}.}
    \label{fig:6302_three_panel}
\end{figure}

Doppler velocities derived from the Fe \RomanNumeralCaps{1} 617.3 nm  (Figure \ref{fig:6173_three_panel}) and 630.2 nm (Figure \ref{fig:6302_three_panel}) lines during the 50s of the simulation. These velocities showed the same initial redshift during the period of beam heating, followed by a blueshift when the beam heating stopped. The smaller redshift velocities for both of these lines, compared to Fe \RomanNumeralCaps{1} 630.1 nm, were a result of the  smaller chromospheric contribution to each of these lines (15\% and 14\% respectively). Both lines still show a maximum redshift several seconds after the maximum beam heating. 

\bibliography{paper}{}
\bibliographystyle{aasjournal}

\end{document}